# Spectral imaging at high-definition and high speed in the mid-infrared

David Knez[1], Benjamin W. Toulson[1], Anabel Chen[1], Martin H. Ettenberg[2], Hai Nguyen[2], Eric O. Potma[1], Dmitry A. Fishman[1]

[1]*Department of Chemistry, University of California Irvine, Irvine, CA 92697, USA*

[2]*Princeton Infrared Technologies, Inc., 7 Deerpark Dr. Suite E, Monmouth Junction, NJ 08852, USA*

**Abstract**

Spectral imaging in the mid-infrared (MIR) range provides simultaneous morphological and chemical information of a wide variety of samples. However, current MIR technologies struggle to produce high-definition images over a broad spectral range at acquisition rates that are compatible with real-time processes. We present a novel spectral imaging technique based on non-degenerate two-photon absorption of temporally chirped optical MIR pulses. This new approach avoids complex image processing or reconstruction and enables high-speed acquisition of spectral data cubes [xyω] at high pixel density in under a second.

**Introduction**

In spectral imaging, each pixel in a two-dimensional [xy] image is expanded along the optical frequency [ω] axis, yielding an information-rich data cube [xyω] of both spatial and spectral information. When realized in the mid-infrared (MIR) spectral region, spectral imaging allows direct spatial differentiation of chemical composition – providing an indispensable tool for many applications in chemical, medical and bio-related fields. MIR spectral imaging has been demonstrated to great effect for label-free histopathological imaging [1,2], stand-off detection of materials [3-6], gas analysis [7], and environmental surveying [8]. Yet, a broader utilization of the technique for applications that require both speed and high-definition imaging capabilities has remained challenging. Conventional interferometric Fourier-transform infrared (FTIR) methods are generally too slow for rapid analysis or *in situ* observations of real-time processes[9]. In this context, there is a need for MIR spectral imaging technologies that combine high pixel density mapping with high acquisition and image processing speed.

MIR imaging historically suffers from technological limitations associated with detection of infrared radiation. Traditional detection approaches are based on the linear absorption of MIR light by small bandgap semiconductors (MCT, InSb). Such single element or matrix arrays are susceptible to thermal noise, and thus require active cooling, often down to cryogenic temperatures. In addition, for arrayed detectors there is a trade-off between pixel density and the overall response time of the detector assembly. Whereas readout times of small focal plane arrays (up to 128 x128) can be fast, the frame rate of larger MIR-sensitive arrays is typically too slow for high-speed imaging applications.[9] Moreover, wide-field FTIR imaging uses a low brightness broadband source and necessitates an interferometric scan plus subsequent processing to reconstruct the MIR absorption spectrum, which can slow down the overall acquisition process. Recent approaches based on quantum cascade lasers (QCL) benefit from higher brightness and an alternative spectral scanning scheme that involves rapid sweeping of a narrow spectral window [1,2]. For instance, MIR spectral imaging covering a full spectral octave with four synchronized QCL lasers has been achieved, enabling hyperspectral data cube acquisition within ~3 seconds using a 128x128 pixel focal plane array.[2] Given the limited number of pixels of the array, mosaicking over a larger sample area was needed to collect images of higher definition, extending the overall collection time to several minutes.

A new cohort of MIR imaging approaches has been developed to address the aforementioned limitations of arrayed MIR detectors. These methods make use of a conversion step through which information encoded in MIR part of the spectrum is transferred to the visible or near-IR range of the spectrum, where more mature large bandgap semiconductor technology can be used. Such a frequency conversion step can be realized in a variety of ways. For instance, several approaches employ the nonlinear properties of the sample material itself [10-15]. Other methods use an external nonlinear medium to either reconstruct the image through nonlinear interferometry of entangled photons [16, 17] or, alternatively, via direct up-conversion of mid-IR light through a $\chi^{(2)}$-mediated parametric process [18-20]. In this fashion, MIR images have been collected in as little as 2.5 ms per frame using up-conversion in wide-field imaging mode [21]. Although up-conversion enables fast monochromatic imaging, the need for phase-matching adds complexity in both image reconstruction and spectral data acquisition, which can limit a flexible implementation of spectral imaging over a broad bandwidth.

Another conversion method exploits the intrinsic optical nonlinearity of the detector material itself. This detection principle is based on a $\chi^{(3)}$-mediated nonlinear process, namely non-degenerate two-photon absorption (NTA).[22-24] In our previous work, we have used NTA in large bandgap semiconductor materials to directly detect MIR radiation on the chip of visible or near-IR cameras.[25-27]. This approach allowed us to collect chemically selective MIR images with a mega-pixel InGaAs camera, at 500 Hz frame rates. Whereas NTA enables fast MIR imaging with high definition, this principle has not yet been used for spectral imaging. In the current work we introduce a new spectral scanning strategy that takes advantage of the NTA imaging conditions. Using temporally chirped mid-IR pulses and a short near-IR gate pulse, we show that MIR spectral data [xyω] cubes over a >400 cm$^{-1}$ spectral range can be acquired in under 1 s with a >1Mpx InGaAs camera. Without the necessity for image processing or reconstruction, this approach has the potential for true video-rate hyperspectral data acquisition of live processes, enabling rapid data acquisition in all four dimensions, i.e. a [xyωt] hypercube.

**Results**

Our high-speed spectral imaging technique leverages the broad spectral width of an ultrashort MIR pulse, as illustrated in Figure 1. Consider first a MIR pulse with a temporal width of 40 fs that is bandwidth-limited. In the frequency domain, such a pulse manifests itself with a constant spectral phase, as can be inferred from a simple Fourier transformation (see Supplementary Information). In the time-frequency plot of Figure 1a, the constant spectral phase is associated with a narrow temporal distribution of the different frequency components in the pulse. When the bandwidth-limited pulse traverses a sample, for instance a thin layer of polystyrene, its spectrum is modified due to the linear absorption of light by the polymer (Figure 1a). The pulse spectrum now shows narrow spectral features in its otherwise broad spectral profile. In the time-domain, the narrow imprints in the MIR pulse spectrum produce almost invisible modification to the temporal profile, most notably a decrease of the overall intensity and the introduction of an extremely shallow pedestal, as shown in Figure 1b. The NTA signal is obtained by temporally overlapping the MIR pulse with a short near-IR gate pulse on the detector. If the gate pulse is scanned in time relative to the MIR pulse, the convolution results in a profile that lacks clearly resolved features related to the MIR spectral imprints.

The situation is different if a linear chirp is applied to the MIR pulse. The time-frequency plot in the inset of Figure 1c shows the original MIR pulse that has been linearly chirped to a temporal width of 4.5 ps, corresponding to a quadratic spectral phase (see Supplementary Information). Different frequency components now have their maximum intensity at different moments in time. Whereas the spectral phase has no direct effect on the absorption imprints in the pulse's power spectrum that is observed in the frequency domain (Figure 1c), it has a profound effect on the MIR pulse in the time-domain.

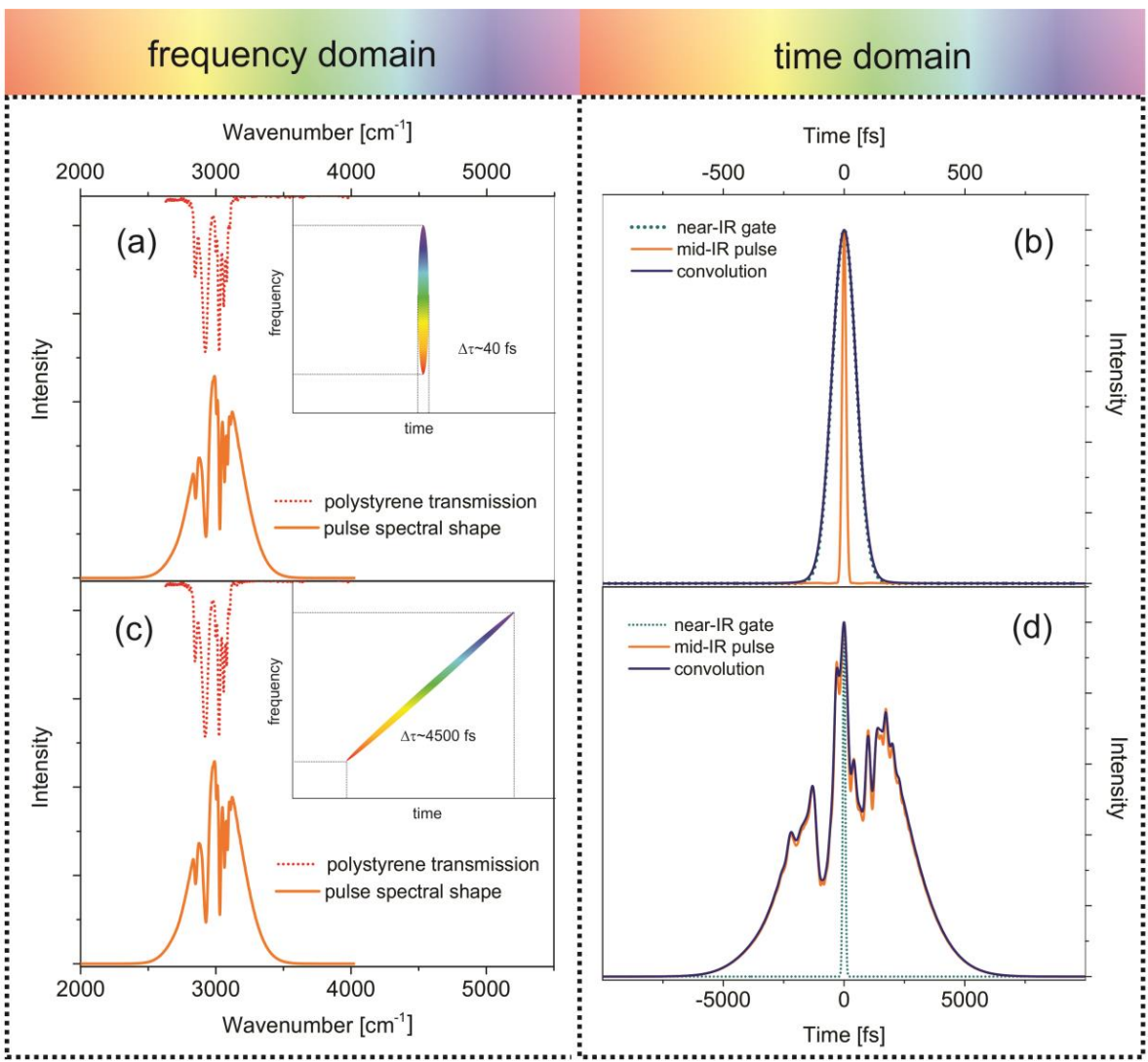

**Figure 1.** Concept of chirped pulse NTA (CP-NTA). The broad pulse spectrum is modified due to linear absorption of light by the sample (a, c). Inset shows a time-frequency plot of the MIR pulse in the absence or presence of linear chirp. (b, d) The resulting pulse shape in the time-domain without (b) and with (d) introduction pulse chirp. Dark blue lines illustrate the convolution between the MIR and gate pulses.

As shown in Figure 1d, the time-resolved MIR pulse shows variations that mimic the spectral features in the frequency domain. In the case of linear chirp, there is a linear relationship between the time and frequency axis, allowing a direct reading of the spectrally encoded information in the time domain. Utilizing the gated principle of NTA detection, the MIR pulse spectrum can be directly accessed through a rapid scan of the MIR and gate relative time delay, i.e. a cross-correlation. Note that for sufficiently broadened MIR pulses, the spectral information is largely unaffected by the convolution with the gate pulse. The use of chirped pulses for encoding spectral information in time-domain signals has been demonstrated for various forms of spectroscopy [28], and has recently been an important approach for achieving high spectral resolution in coherent Raman scattering microscopy with broad bandwidth pulses [29, 30]. Near-IR chirped pulses have also been used to upconvert MIR signals to the visible range with the aid of nonlinear crystals [31, 32]. Here we show that this principle can be extended to MIR spectral imaging through NTA by chirping the MIR pulse instead.

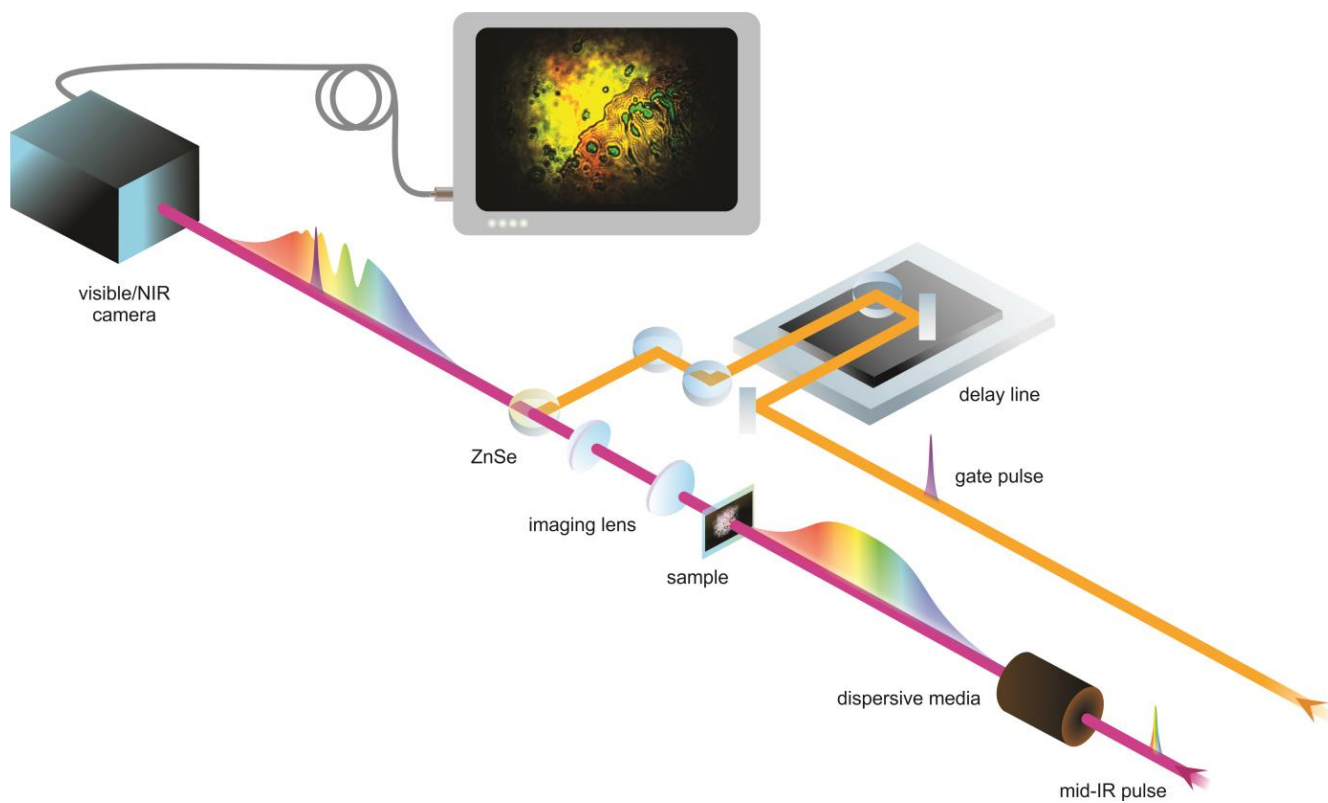

**Figure 2.** Experimental scheme of chirped-pulse NTA (CP-NTA). An ultrashort MIR pulse is chirped by propagation through a 15 cm Si rod and spatially overlapped with a temporally short gate pulse on the camera chip. Control of the MIR-gate pulse time delay permits a rapid spectral scan across the MIR pulse spectrum.

A schematic of the experimental implementation of the chirped pulse non-degenerate two-photon absorption (CP-NTA) imaging technique is shown in Figure 2. We use an ultrashort mid-IR pulse centered at 3.33 µm (3000 cm$^{-1}$) that is stretched by a 15 cm long Si rod (see *Methods and Materials*). The stretched pulse is subsequently passed through the sample and imaged with two CaF$_2$ lenses onto a >1.3 Mpx lattice match In$_{0.53}$Ga$_{0.47}$As camera chip (0.734 eV, 1690 nm bandgap). The near-IR gate pulse is centered at 1900 nm (5263 cm$^{-1}$) with a temporal width of 116 fs. The gate pulse passes through a delay stage before it is co-incident with the MIR pulse on the detector chip. Note that the camera chip is covered with a fused silica protective window, which significantly attenuates the mid-IR beam (0.75 OD @ 3000 cm$^{-1}$, see *Supplementary information*, Figure SF3). More details on the experimental parameters can be found in the Supplementary Information (see Table S1).

The linear relation between the time and frequency axis in CP-NTA relies on the successful application of linear chirp. This can be achieved with MIR transparent materials that exhibit sufficient group velocity dispersion with limited higher-order dispersion. We have found that such can be realized with a Si rod. The MIR pulse spectrum is shown in Figure 3a, indicating a full width at half maximum (FWHM) of 360 cm$^{-1}$, as measured with a grating-based spectrometer equipped with a single pixel MCT detector (~20 seconds per spectral point, see *Methods and Materials*). This spectrum supports a bandwidth-limited temporal width of 40 fs. Figure 3b depicts the NTA signal on the camera as a function of the MIR-gate delay time. We can find the temporal width of the MIR pulse from the NTA cross correlation of the MIR and gate pulses. In the absence of the Si rod, the MIR pulse duration is found as 160 fs, which is longer than the minimum width of 40 fs because of residual dispersion by the optical elements in the setup. Upon insertion of the Si material in the MIR beam, the NTA cross correlation is significantly elongated in time, revealing a MIR pulse with a temporal width of 4.5 ps.

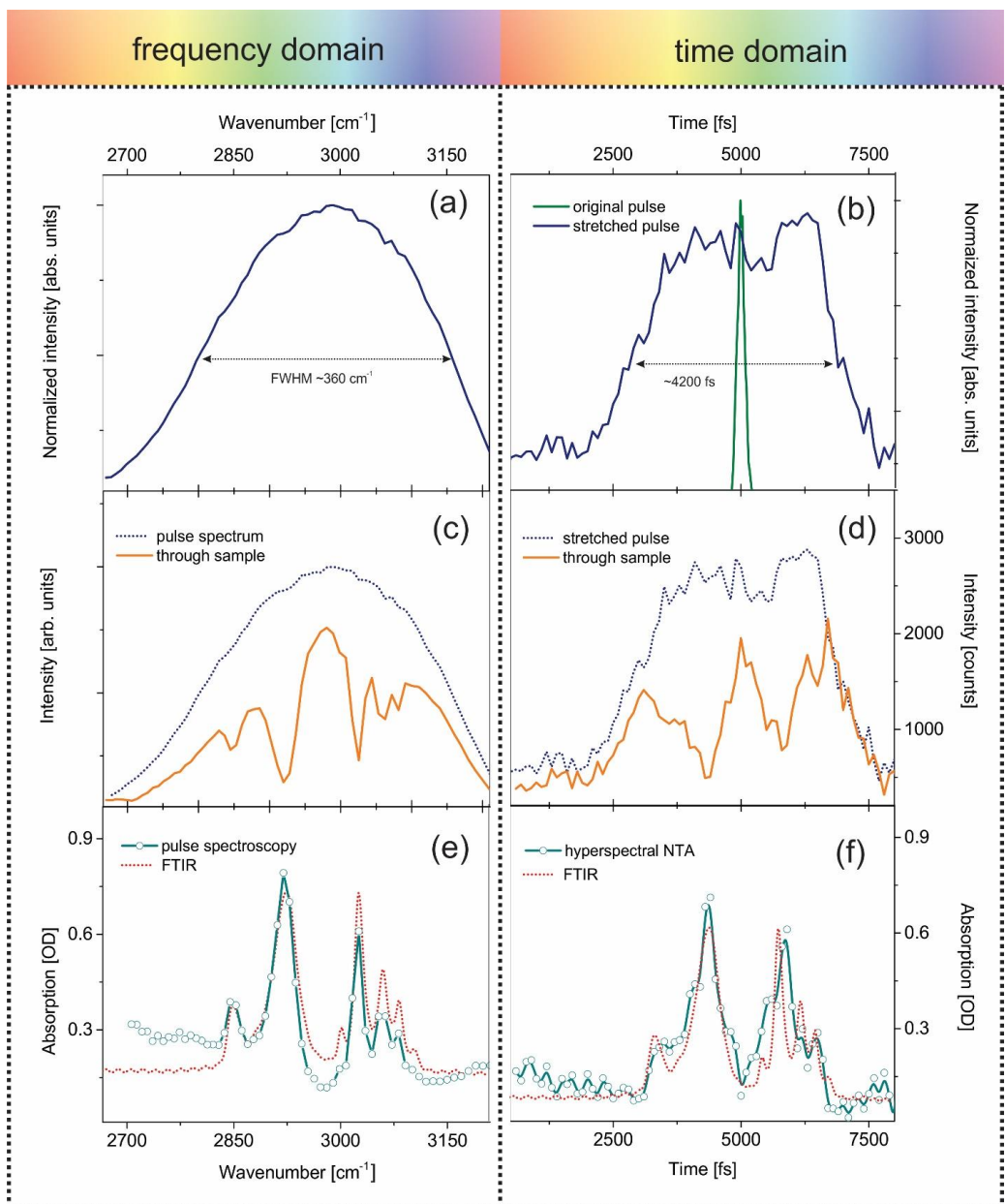

**Figure 3.** Side-by-side demonstration of spectroscopy on polymer standard (7 μm polystyrene) in frequency-domain (spectrometer, left panels, a,c) and time-domain (CP-NTA, right panels b,d). (e,f) Retrieved absorption spectra are in good agreement with the polystyrene absorption lines.

When the MIR pulse passes through a 7 μm layer of polystyrene, a frequency domain measurement reveals a pulse spectrum with narrowband imprints of vibrational absorption lines, as is evident from Figure 3c. The absorption features correspond to the symmetric and asymmetric stretching vibrations of the carbon-hydrogen bonds of polystyrene. Using the chirped pulse, this spectral information can also be obtained in the time-domain, as shown by the NTA cross correlation in Figure 3d. The time-domain measurement was collected with a 16 ms dwell time per spectral bin for a total effective acquisition time of 800 ms. In the current configuration, the available spectral range can be extended to ~530 cm$^{-1}$ for a 1.1 s effective acquisition time per spectral sweep.

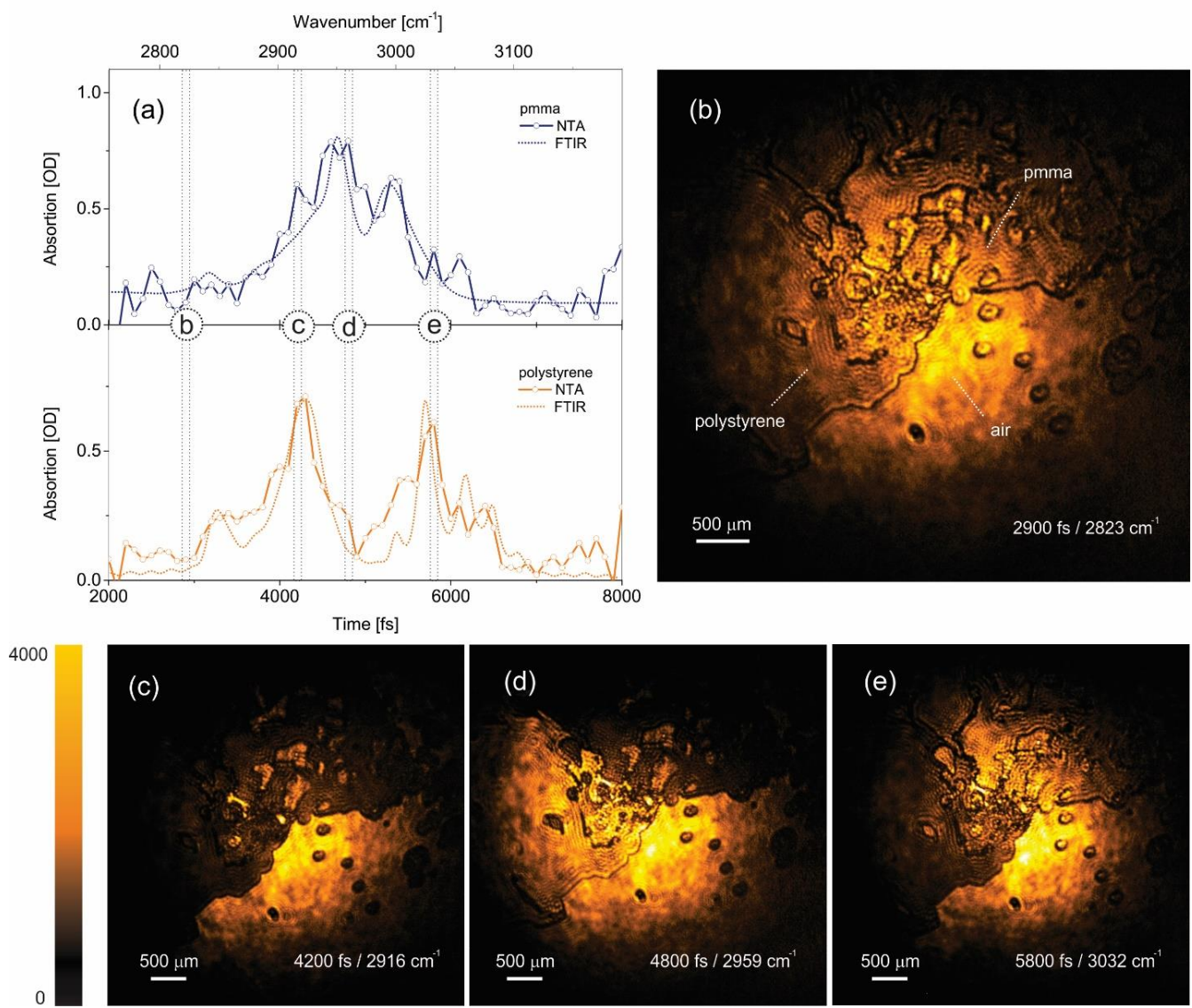

**Figure 4.** Spectral imaging of a polymethyl methacrylate / polystyrene film. (a) Polymer absorption spectrum extracted from the hyperspectral data cube. (b, c, d, e) Raw transmission images at different positions in time/spectral domain, directly revealing spatially dependent chemical information in the field of view. Acquisition time is 16 ms per frame/spectral point.

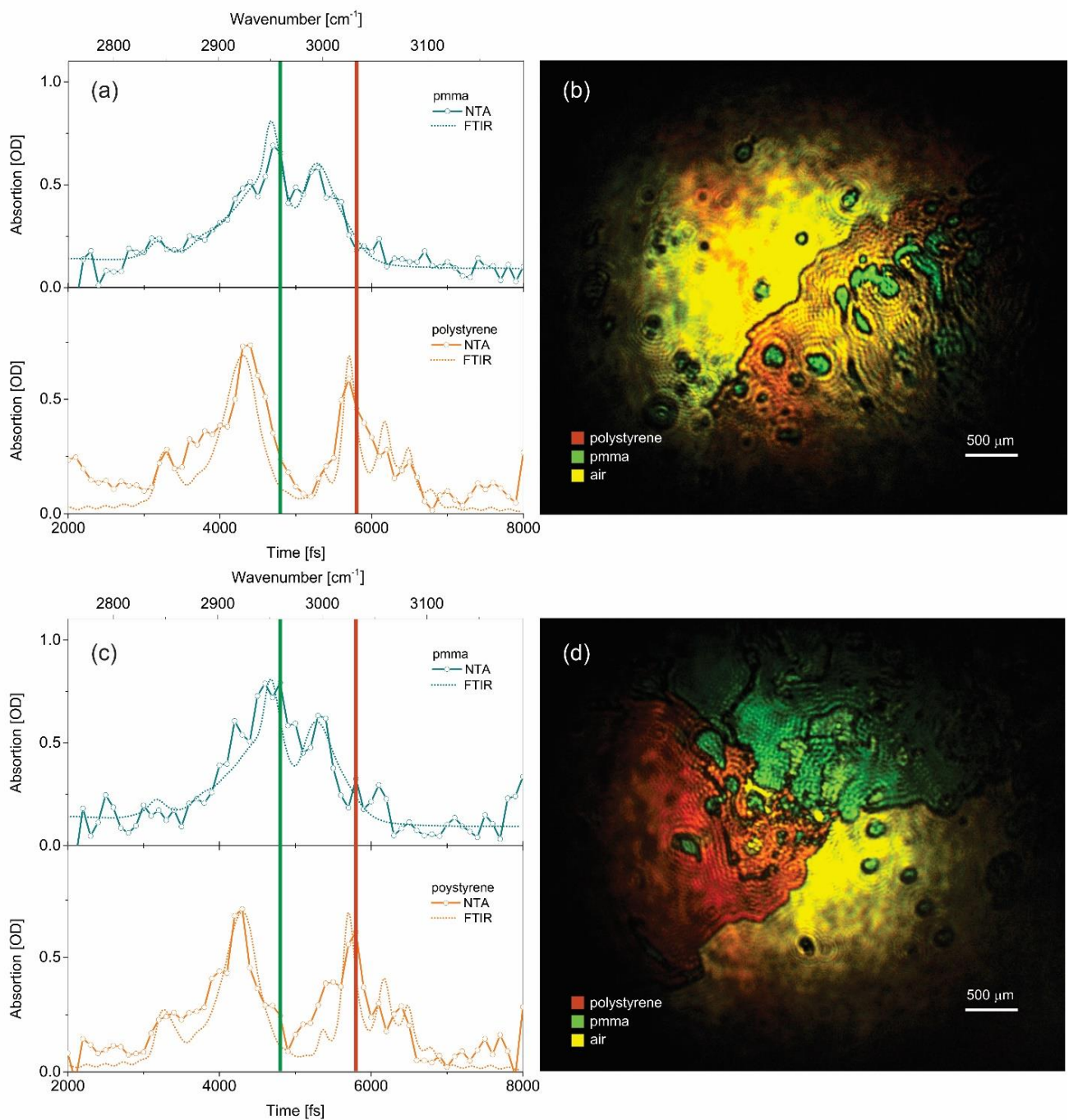

**Figure 5**. Spectral imaging of polymethyl methacrylate/polystyrene films. Dimensionality of the hyperspectral data stack is reduced through overlay of red and green color channels measured on and off the absorption resonance of a given polymer. Effective image acquisition time is ~32 ms (16 ms per frame).

The corresponding MIR absorption spectrum can be derived from the measured pulse spectra, either in the frequency-domain (Figure 3e) or in the time-domain (Figure 3f). Both data sets show good agreement with the signatures in the FTIR spectrum of polystyrene. Using the positions of the polystyrene absorption lines as calibration standards, we determine a chirp rate of $\beta \sim 0.072$ cm$^{-1}$/fs (see Supplementary Information). This corresponds to a ~8.4 cm$^{-1}$ instrument response function, which spans the frequency window of temporal overlap between the short gate pulse and the chirped MIR pulse. The latter is also confirmed from an analysis of the main polystyrene lines retrieved from CP-NTA. The spectral features at 2923 cm$^{-1}$ (FTIR FWHM=35 cm$^{-1}$) and 3025 cm$^{-1}$ (FTIR FWHM=15 cm$^{-1}$) yield spectral widths of 38 cm$^{-1}$ and 19 cm$^{-1}$, respectively, after convolution with the 8.4 cm$^{-1}$ resolution function (see *Supplementary Information*, Figure SF4).

The linearity of the temporal chirp of the MIR pulse is important for direct frequency-to-time conversion. To confirm the linear distribution of frequency components, i.e. the linearity of the instantaneous frequency, we perform CP-NTA spectroscopic imaging with pulses of various carrier frequencies $\omega_0$ around the polystyrene absorption bands (*Supplementary Information*, Figure SF5). The data shows that the line positions and overall absorption spectra remain identical regardless of carrier frequency of the MIR pulse. This observation indicates good linearity of the instantaneous frequency distribution across the full width of the MIR pulse.

Figure 4 shows an example of unprocessed hyperspectral imaging data for a sample composed of adjacent polymethyl methacrylate (PMMA) and polystyrene layers. These materials feature several distinct absorption lines near 2900 cm$^{-1}$. When the sampled MIR frequency is tuned to 2823 cm$^{-1}$ (2900 fs in time domain), which is off-resonance for both compounds, both layers appear transparent (point b on Figure 4a and Figure 4b). Upon rapidly scanning the time delay, the polymers appear darker or brighter, following their absorption spectral profile. Figures 4c-e show several unprocessed frames from the hyperspectral data cube, revealing clear transmission differences between the different materials due to their absorption profiles.

To reduce the dimensionality of the hyperspectral data stack and demonstrate chemical differentiation, we select two images from the data cube where one polymer appears transparent, while another exhibits strong absorption and vice versa (green and red spectral points, Figure 5a and 5c). These two frames are represented as the red [R=100, G=0, B=0] and green [R=0,G=100,B=0] channels and color-merged to form the single image. The results are shown in Figures 5(b) and 5(d), which clearly differentiate the spatial distribution of the two polymers within the field of view. Similarly, Figure 6 displays reduced dimensionality spectral images of various organic materials and their combinations images though CP-NTA, including (a) polystyrene, (b, f) a polymethyl methacrylate/polystyrene structure, (c) ethanol, (d) silicone lubricant (polydimethylsiloxane as main component), and (e) a melted polyethylene flake. The images comprise of two 16 ms frames at two different spectral positions, separated by 72 cm$^{-1}$ in the frequency domain (1 ps separation in time), and represent the on and off resonance positions of a given material. Overlay of the red and green frames then yields an image with chemical contrast.

**Discussion**

In this work we have introduced a new MIR spectral imaging approach, namely chirped-pulse nondegenerate two-photon absorption, or CP-NTA for short. This technique takes advantage of the attractive features NTA-based MIR detection. Chief among these advantages is the ability to collect MIR images at high definition. Unlike the relatively low pixel density matrix of the fastest focal plane arrays, the >1 Mpx detector chips used in NTA offer much improved sampling of the projected MIR image. This high-definition imaging capability does not come at the expense of speed, as NTA benefits from the mature readout technology of visible/near-IR cameras, permitting MIR imaging at high frame rates. In addition, the NTA technique is interferometry free and does not rely on phase matching of the MIR and gate pulses, enabling a robust and relatively simple optical setup.[26, 27] Furthermore, NTA does not require any image reconstruction or processing for immediate visualization, allowing direct and on-screen observation of fast processes at particular MIR frequencies.

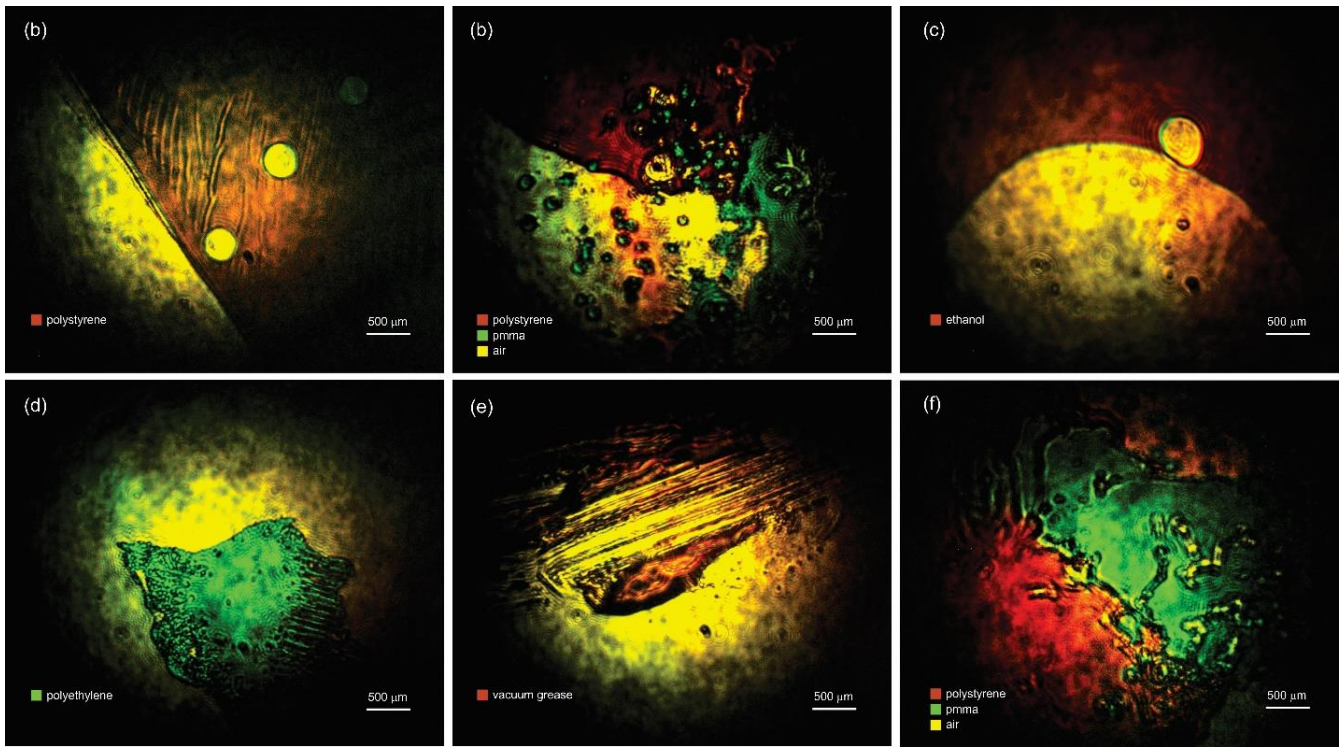

**Figure 6**. Hyperspectral imaging of various materials and material combinations – (a) polystyrene, (b, f) polymethyl methacrylate / polystyrene hybrids, (c) – ethanol, (d) – polyethylene, (e) – silicone lubricant (polydimethylsiloxane). Effective image acquisition time 32 ms (16 ms per frame).

The key advance in this work is the expansion of NTA-based MIR imaging along the frequency dimension. The CP-NTA method utilizes the inherent gated detection property of the technique and achieves spectral imaging through time-domain detection by simply adding a dispersive material in the MIR beam. Spectral tuning requires scanning a delay line, which can be performed with ease, at high speed, and avoids the need for detuning a light source.

In CP-NTA each spectral image is acquired in wide-field mode, collecting spectral data at each pixel in a massively parallel fashion. In contrast to laser-scanning approaches or wide-field imaging with low pixel density arrays, which require mosaic stitching to achieve high-definition images, the CP-NTA approach records frames of high pixel number by using >1.3Mpx camera chips. In this work, we have used a 16 ms image acquisition time for a ~500 kpx frame and a >700 px MIR spot size. This camera frame rate (62.5 Hz, 16 ms) has been chosen to eliminate signal fluctuations while allowing full matrix array readout. We note that there is room for significant improvement of the CP-NTA imaging conditions. For instance, the current CP-NTA signals have been collected without synchronization of the camera to the 1 kHz radiation source, causing additional noise at frame rates beyond 70 Hz. With appropriate electronic synchronization and a more compact source-detection arrangement, the MIR acquisition rate for the used camera can be easily increased to 100 Hz for the full 1280x1024 window.

In the spectral domain, the time-frequency sweep can be carried out in a variety of ways. In this work we have used spectral imaging over the full spectral range of the MIR pulse, but we have also performed MIR imaging with reduced spectral sampling. For example, the data representation in Figures 5 and 6 is based on only two spectral frames with a fixed time-frequency separation. The 72 cm$^{-1}$ spectral separation between these two images (1 ps in time) can be acquired within a few milliseconds or faster. Rapid modulation of the spectral position can for instance be achieved by chopping the gate pulse beam with 0.7 mm glass windows at rates exceeding 100 Hz, producing on and off frames for high-speed chemical imaging.

In the present experiments we have utilized a motorized linear stage that allows a fine spectral sweep with scanning speeds up to 12,000 cm$^{-1}$/s with 0.72 cm$^{-1}$ precision (~170000 fs/s, 10 fs). Other inexpensive approaches exist for rapidly controlling the time-delay setting and performing rapid delay modulation over ~1.5 mm (10 ps) distances. For example, commercially available travel delay stages allow >20 Hz delay modulation over an extended 15 ps range with 10 fs time delay precision. Some of the concepts used for the recent development of rapid delay lines in coherent Raman scattering [33-35] can also be translated for use in CP-NTA. We are confident that further development of rapid scanning of the time-frequency axis will render CP-NTA suitable for video-rate hyperspectral data acquisition, in which full [xyω] data cubes can be collected in real-time.

## Materials and Methods
### Sample preparation

The polymer samples (polystyrene and polymethyl methacrylate) are placed between two coverslips (170 μm) and melted using a hotplate. During the melting process, the polymer sample is pressed flat by a weight until the desired thickness is reached. The methanol sample is prepared by applying 10 μL of the liquid between two coverslips. The sample is subsequently agitated until bubbles appear. Polyethylene samples are cut and placed between two coverslips. Vacuum grease samples are prepared by smearing a small amount of the substance onto a coverslip.

### FTIR experiments

Conventional infrared absorption spectra are measured using a Jasco 4700 FTIR spectrometer in attenuated total reflection (ATR) geometries. For the ATR experiments, the Jasco ATR-Pro One accessory equipped with a diamond crystal is used. The spectra are averaged over 40 scans and are acquired with a 2 cm$^{-1}$ resolution.

### Frequency domain experiments with pulsed radiation

Frequency domain experiments for pulse characterization and spectroscopy are performed using a grating-based spectrometer (CM110, Spectral Products, 150 grooves/mm) equipped with single pixel amplified MCT detector (PDAVJ10, Thorlabs). All spectra have been measured with 10 nm spectral resolution (~11 cm$^{-1}$).

### CP-NTA detection

A 1 kHz amplified femtosecond laser system (Spitfire Ace, Spectra Physics) is used to seed two optical parametric amplifiers (OPA, Topas Prime, Light Conversion). One OPA is used as a source of NIR gate radiation at 1900 nm (0.65 eV). The signal and idler pulses from the second OPA system are used to generate mid-IR pulses through the process of difference frequency generation in a nonlinear medium. Both mid-IR and near-IR pulses are recombined on a 1 mm thick ZnSe window, which serves as dichroic mirror, after which the pulses are overlapped on an InGaAs camera chip (1280MVCam, Princeton Infrared Technologies Inc.). Temporal overlap is controlled through an automated delay stage (GTS150, Newport). The MIR imaging system consists of two 100 mm CaF$_2$ lenses, resulting in a 1:1 imaging system with NA = 0.015. All experiments are performed with a fixed NIR irradiance of 240 MW/cm$^2$ and MIR irradiance 0.25 MW/cm$^2$ on the camera chip.

**Acknowledgements**
DAF thanks Yulia Davydova for all the help and support. EOP acknowledges NIH R01GM132506. EOP and DAF acknowledge NIH R21GM141774. We thank the Laser Spectroscopy Labs. DAF wants to dedicate this work to dear friend Luke.